\UseRawInputEncoding 
\documentclass[11pt,a4paper]{article}
\usepackage{amsfonts}
\usepackage{amssymb}
\usepackage{mathrsfs}
\usepackage{amsmath}
\usepackage{booktabs}
\usepackage{epsf,epsfig,amsfonts,amsgen,indentfirst}
\usepackage{amsmath,amstext,amsbsy,amsopn,amsthm,bbding,wasysym}
\usepackage{multicol,mathdots}
\usepackage{subfigure}
\usepackage[numbers,sort&compress]{natbib}
\allowdisplaybreaks

\newtheorem{theorem}{Theorem}[section]

\newtheorem{lemma}[theorem]{Lemma}

\begin{document}
\setlength{\emergencystretch}{5em}
\title
{\LARGE \textbf{Near-synchronous multiphoton quantum interference
\thanks{Supported by NSFC (No. 12261071) and NSF of Qinghai Province (No. 2025-ZJ-902T).} }}

\author{Xiangshuai Dong$^a$, Tingzeng Wu$^{b,c}$\thanks{{Corresponding author.\newline
\emph{E-mail addresses}: mathtzwu@163.com, a3566293588@163.com
}} \\
{\small $^{a}$ School of Mathematical Sciences, Xiamen University,  }\\
{\small  Xiamen, 361005,  P.R.~China} \\
{\small $^{b}$ School of Mathematics and Statistics, Qinghai Minzu University, }\\
{\small  Xining, Qinghai 810007, P.R.~China} \\
{\small $^{c}$ Qinghai Institute of Applied Mathematics, Xining, Qinghai 810007, P.R.~China}}
\date{}

\maketitle
\noindent {\bf Abstract:}
Nonclassical multiphoton interference is an important resource for optical quantum information processing, and its output statistics depend on the distinguishability of the input photons. In this paper, using time delay as a tunable parameter, we theoretically analyze the interference behavior of independent pure-state single photons with identical Gaussian spectral envelopes under near-synchronous conditions. Using the immanant representation of the single-photon internal-state overlap matrix to express the weights of the exchange-symmetry components, we determine the leading asymptotic behavior and approximation error of each weight, and characterize the weight suppression laws when different groups of photons become synchronized on different scales. Given the local identity classes of the delay functions and the vanishing orders of pairwise delay differences, the vanishing orders of the weights of the exchange-symmetry components can be computed in polynomial time by integer minimum-cost flow. On this basis, we provide a network-independent total-variation error bound for the complete output photon-number distribution, establish a quantitative relationship among the photon spectral width, synchronization precision, and interference statistics deviation, and provide a quantitative basis for the analysis and approximation of near-synchronous multiphoton interference.


\section{Introduction}\label{sec1}
Nonclassical multiphoton interference is an important resource for optical quantum information processing. Time delay tunes photon distinguishability by changing wave-packet overlap and thus directly affects interference patterns.
In the Hong--Ou--Mandel (HOM) experiment, when two indistinguishable photons enter different input ports of a balanced beam splitter simultaneously, the coincidence probability between different output ports is zero.
After introducing a relative delay, the coincidence count recovers as the wave-packet overlap decreases, see \cite{hom}.

For more photons, Tillmann et al. \cite{till} decomposed the coincidence probability according to exchange symmetry, described the network contribution in multiphoton transition amplitudes by the immanant of the transition matrix, and separated it from the overlaps between the quantum states associated with the internal degrees of freedom of the input photons (hereinafter referred to as internal states).
Related studies include three-photon interference \cite{tan}, partial distinguishability theory \cite{shch,tichy}, and permutation symmetry of multimode counting \cite{khalid}. In the Gaussian delay model, Spivak et al. \cite{spivak} determined the vanishing conditions of each component when the delays coincide. Related tensor criteria can be found in \cite{berget} and \cite{gamas}.

This paper studies the weights of exchange-symmetry components of near-synchronous multiphoton states and their influence on output statistics. The input consists of independent pure-state single photons from different channels that are identical in their internal states except for time delays and share a common Gaussian spectral envelope.
We adopt an idealized lossless passive linear optical network model whose spatial transformation is described by a frequency-independent unitary matrix. Detectors perform photon-number measurements over the entire wave packet and do not resolve internal degrees of freedom.
Using the immanant of the internal-state Gram matrix, we determine the vanishing orders and leading coefficients of the weights of the exchange-symmetry components as the temporal mismatch decreases, reduce the computation of the vanishing orders to integer minimum-cost flow, and bound the approximation error of the complete output distribution by the total weight of the neglected components.

Let $n$ be a positive integer. A partition of $n$ is a sequence of positive integers
$\lambda=(\lambda_1,\ldots,\lambda_\ell)$ satisfying
$\lambda_1\geq\lambda_2\geq\cdots\geq\lambda_\ell>0$ and
$\sum_{r=1}^{\ell}\lambda_r=n$, denoted by $\lambda\vdash n$.
We set $\lambda_r=0$ for $r>\ell$.
The Young diagram of $\lambda$ consists of $n$ boxes aligned to the left, with the $r$-th row containing $\lambda_r$ boxes.
Interchanging the rows and columns of this diagram gives the Young diagram of the conjugate partition $\lambda'$, where
$\lambda'_j=\#\{r:\lambda_r\geq j\}$ is the length of the $j$-th column of the original diagram. Let $S_n$ be the symmetric group on $\{1,\ldots,n\}$ and $e$ its identity element.
The complex irreducible characters of $S_n$ are labeled by partitions of $n$.
For $\lambda\vdash n$, let $\chi_{\lambda}$ be the irreducible character corresponding to $\lambda$,
and $d_\lambda=\chi_{\lambda}(e)$ its degree, i.e., the dimension of the corresponding irreducible representation.
For an $n\times n$ matrix $B=[b_{ij}]$, the generalized matrix function
\emph{immanant} associated with $\chi_{\lambda}$ is defined as
\begin{equation}
 \operatorname{Imm}_\lambda(B)=\sum_{\sigma\in S_n}\chi_{\lambda}(\sigma)
                         \prod_{i=1}^n b_{i\sigma(i)}.
 \label{eq1}
\end{equation}
For $\lambda=(n)$, $\operatorname{Imm}_{(n)}(B)$ is the permanent. For $\lambda=(1^n)$, $\operatorname{Imm}_{(1^n)}(B)$ is the determinant.

Let $s>0$ be the angular-frequency standard deviation of the common spectral probability density,
and let $\theta_i$ be the time delay of the $i$-th photon.
Denote by $y_i=s\theta_i$ the dimensionless delay, and let $\boldsymbol{y}=(y_1,\ldots,y_n)$.
The pairwise inner products of the internal states of the photons form the Gram matrix $G(\boldsymbol{y})$.
For a common Gaussian spectral envelope, this matrix and the corresponding exchange-symmetry component weights are
\[
 G(\boldsymbol{y})_{ij}=e^{-(y_i-y_j)^2/2},\qquad
 p_\lambda(\boldsymbol{y})=\frac{d_\lambda}{n!}\operatorname{Imm}_\lambda(G(\boldsymbol{y})).
\]
The diagonal entries of $G(\boldsymbol{y})$ are $1$, and the off-diagonal entries describe the internal-state overlap of the corresponding two photons.
The partition $\lambda$ labels the irreducible representation type under photon exchange,
and $p_\lambda(\boldsymbol{y})$ is the total weight of all components of type $\lambda$ in the spatially reduced state, satisfying
\[
 p_\lambda(\boldsymbol{y})\geq0,\qquad
 \sum_{\lambda\vdash n}p_\lambda(\boldsymbol{y})=1.
\]
These weights are determined by the internal-state overlaps of the input photons and characterize the exchange-symmetry components of the input state. The probability of a given output photon-number configuration also depends on the normalized spatial states of the exchange-symmetry components and the transition amplitudes of the optical network. Under perfect synchronization, all delays are equal, all entries of $G(\boldsymbol{y})$ are $1$, hence $p_{(n)}(\boldsymbol{y})=1$ and the weights of all other types vanish. This paper characterizes how the other components are suppressed as the temporal mismatch decreases.

Filling the Young diagram of $\lambda$ with each of $1,\ldots,n$ used exactly once gives a tableau $T$ of shape $\lambda$.
Let $\mathcal{T}_\lambda$ denote the set of all such tableaux.
Here the entries are not required to increase along rows or columns, so $\mathcal{T}_\lambda$ contains $n!$ tableaux in total.
Let $T(r,j)$ denote the label in row $r$ and column $j$. Define
\[
 b_\lambda=\sum_r(r-1)\lambda_r=\sum_j\binom{\lambda'_j}{2},\qquad
 c_\lambda=\prod_j\lambda'_j!,\qquad
 q_\lambda=\prod_r((r-1)!)^{\lambda_r}.
\]
Here the sums and products run over the rows or columns of the Young diagram.
$b_\lambda$ is the total number of pairs of distinct boxes within columns,
$c_\lambda$ is the product of the numbers of internal arrangements in the columns, and $q_\lambda$ is the factorial factor determined by the row lengths.
For $\boldsymbol{x}=(x_1,\ldots,x_n)\in\mathbb{R}^n$ and $T\in\mathcal{T}_\lambda$, define the column-difference product
\[
 \Delta_T(\boldsymbol{x})=\prod_j\prod_{1\leq r<u\leq\lambda'_j}
                  (x_{T(u,j)}-x_{T(r,j)}).
\]
In each column, this product takes the differences between the coordinates corresponding to labels in lower boxes and those corresponding to labels in upper boxes, and then multiplies over all columns; hence it is a homogeneous polynomial of degree $b_\lambda$. The empty product is set to $1$.
Furthermore, let
\begin{equation}
 \begin{aligned}
 \bar y&=\frac1n\sum_i y_i,\qquad
 S(\boldsymbol{y})=\sum_i(y_i-\bar y)^2,\\
 \kappa_\lambda(\boldsymbol{y})&=\frac{d_\lambda}{n!c_\lambda q_\lambda}
                    \sum_{T\in\mathcal{T}_\lambda}\Delta_T(\boldsymbol{y})^2.
 \end{aligned}
 \label{eq2}
\end{equation}
Here $\bar y$ is the average dimensionless delay, and $S(\boldsymbol{y})$ is the dimensionless sum of squares of the temporal mismatch of each photon relative to the average delay.
It equals zero when all delays are equal and tends to zero when all delays tend to the same value.
$\kappa_\lambda(\boldsymbol{y})$ is a nonnegative homogeneous polynomial of degree $2b_\lambda$ formed by the squares of the column-difference products over all tableaux; we will show below that it gives the leading-order approximation of $p_\lambda(\boldsymbol{y})$ near synchronization.
For $\lambda,\nu\vdash n$, if
$\sum_{r=1}^k\lambda_r\geq\sum_{r=1}^k\nu_r$ holds for every $k\geq1$,
then $\lambda$ is said to dominate $\nu$, written $\lambda\succeq\nu$;
in comparisons, zeros are appended to the ends of the partitions. In this paper, asymptotic expansions are taken as $\varepsilon\to0$.

\begin{theorem}\label{thm1}
Let $\lambda\vdash n$ and $\boldsymbol{y}\in\mathbb{R}^n$. Then
\begin{equation}
 e^{-S(\boldsymbol{y})}\kappa_\lambda(\boldsymbol{y})
 \leq p_\lambda(\boldsymbol{y})\leq\kappa_\lambda(\boldsymbol{y}).
 \label{eq3}
\end{equation}
Let $\boldsymbol{x}\in\mathbb{R}^n$, and let $\nu$ be the multiplicity partition of its equal coordinates. If $\lambda\not\succeq\nu$, then $p_\lambda(\varepsilon\boldsymbol{x})\equiv0$; if $\lambda\succeq\nu$, then $\kappa_\lambda(\boldsymbol{x})>0$, and
\[
 p_\lambda(\varepsilon\boldsymbol{x})
 =\kappa_\lambda(\boldsymbol{x})\varepsilon^{2b_\lambda}
       +O(|\varepsilon|^{2b_\lambda+2}).
\]
For any $\boldsymbol{x}\in\mathbb{R}^n$ and $\varepsilon\in\mathbb{R}$,
\[
 0\leq\kappa_\lambda(\boldsymbol{x})\varepsilon^{2b_\lambda}
          -p_\lambda(\varepsilon\boldsymbol{x})
 \leq S(\boldsymbol{x})\kappa_\lambda(\boldsymbol{x})|\varepsilon|^{2b_\lambda+2}.
\]
\end{theorem}

Theorem~\ref{thm1} gives the power laws and approximation errors of the exchange-symmetry component weights allowed by the dominance condition as the delay scale varies. In experiments, one may first improve the temporal matching of photons within groups and then adjust the delays between groups to synchronize the groups.
In this case, the delay variation scales within and between groups are different, and the pairwise delay differences must be recorded.

Let $y_1(\varepsilon),\ldots,y_n(\varepsilon)$ be real-analytic delay functions near zero, and abbreviate $p_\lambda(\varepsilon)=p_\lambda(\boldsymbol{y}(\varepsilon))$. For delay differences that are not identically zero, let
\[
 y_k(\varepsilon)-y_i(\varepsilon)
 =a_{ik}\varepsilon^{v_{ik}}+O(|\varepsilon|^{v_{ik}+1}),
 \qquad v_{ik}\in\mathbb{N}_0,\quad a_{ik}\ne0\quad(i<k).
\]
Here $v_{ik}$ is the lowest nonzero Taylor order. If two functions coincide in a neighborhood of zero, they are said to be locally identical, and we set $v_{ik}=+\infty$. We adopt the conventions $v_{ki}=v_{ik}$ and $a_{ki}=-a_{ik}$. For any $T\in\mathcal{T}_\lambda$, define
\begin{equation}
 w(T)=\sum_j\sum_{1\leq r<u\leq\lambda'_j}
              v_{T(r,j),T(u,j)},\qquad
 H_\lambda=\min_{T\in\mathcal{T}_\lambda}w(T).
 \label{eq4}
\end{equation}
When $w(T)<\infty$, let $A_T=\prod_j\prod_{r<u}a_{T(r,j),T(u,j)}$. The empty sum is zero and the empty product is one. A sum containing an infinite term is infinite. The lowest nonzero Taylor order of a real-analytic function that is not identically zero is called its vanishing order and is denoted by $\operatorname{ord}_0$. Order zero means that the function is nonzero at zero; the order of the identically zero function is set to $+\infty$.

\begin{theorem}\label{thm2}
Let $\lambda\vdash n$, let $y_1,\ldots,y_n$ be real-analytic functions near zero, and let $H_\lambda$ be defined by \eqref{eq4}. If $H_\lambda<\infty$, then
\[
 p_\lambda(\varepsilon)=C_\lambda\varepsilon^{2H_\lambda}
              +O(|\varepsilon|^{2H_\lambda+1}),\qquad C_\lambda>0.
\]
In this case, define
\[
 K_\lambda=\frac{d_\lambda}{n!c_\lambda q_\lambda}
           \sum_{T:\,w(T)=H_\lambda}A_T^2.
\]
Then $e^{-S(\boldsymbol{y}(0))}K_\lambda\leq C_\lambda\leq K_\lambda$. In particular, if all $y_i(0)$ are equal, then $C_\lambda=K_\lambda$. If $H_\lambda=+\infty$, then $p_\lambda$ vanishes identically near zero. Arrange the sizes of the local identity classes in decreasing order as $\nu$; then $H_\lambda<\infty$ if and only if $\lambda\succeq\nu$.
\end{theorem}

Near synchronization, higher-order exchange-symmetry components can be neglected.
Below, the error of the complete output distribution is controlled by the total weight of the neglected components, directly relating synchronization precision to output statistics.

\begin{theorem}\label{thm3}
Let $n$ independent pure-state single photons with identical Gaussian spectral envelopes enter through mutually distinct channels and have identical internal states except for arrival times. Let $\boldsymbol{y}\in\mathbb{R}^n$ be the dimensionless delays, $S=S(\boldsymbol{y})$, and $\rho_{\rm ext}$ the spatially reduced state. Let $E_\lambda^{\rm ext}$ be the orthogonal projection onto the subspace of type $\lambda$. For an integer $K\geq1$, define
\[
 \delta_K=\sum_{\lambda:\,b_\lambda\geq K}p_\lambda(\boldsymbol{y}),
 \qquad P_{<K}=\sum_{\lambda:\,b_\lambda<K}E_\lambda^{\rm ext}.
\]
Then
\begin{equation}
 \delta_K\leq e^{-S}\sum_{m=K}^\infty\frac{S^m}{m!}
 =1-e^{-S}\sum_{m=0}^{K-1}\frac{S^m}{m!}
 \leq\frac{S^K}{K!}.
 \label{eq5}
\end{equation}
Moreover, $1-\delta_K\geq p_{(n)}(\boldsymbol{y})\geq e^{-S}>0$.
Let $\widetilde\rho=P_{<K}\rho_{\rm ext}P_{<K}/(1-\delta_K)$.
After the original spatially reduced state and the normalized truncated state pass through the same lossless passive linear optical network that does not depend on the internal states, their complete output photon-number distributions $P$ and $\widetilde P$ that do not resolve internal states (including multiphoton events within the same output channel) satisfy
\begin{equation}
 \operatorname{TV}(P,\widetilde P)
 :=\frac12\sum_o|P(o)-\widetilde P(o)|
 \leq\frac12\|\rho_{\rm ext}-\widetilde\rho\|_1
 =\delta_K\leq\frac{S^K}{K!},
 \label{eq6}
\end{equation}
where $\|\cdot\|_1$ is the trace norm.
\end{theorem}

The Poisson tail bound in \eqref{eq5} comes from an orthonormal-basis expansion in an auxiliary internal-state space and does not mean that the output photon counts obey a Poisson distribution. The error bound concerns the complete output photon-number distribution. If one renormalizes by conditioning on collision-free events (i.e., events in which at most one photon is detected in each output channel), the probability of such events must also be included. The computation of the weight vanishing orders does not require enumerating all tableaux in \eqref{eq4}.
Using the hierarchical structure of the delay-difference orders, the column assignment of photon labels can be converted into an integer minimum-cost flow, thereby obtaining $H_\lambda$ after the local identity classes and the inter-class orders are given.

\begin{theorem}\label{thm4}
Let $\lambda\vdash n$, and let $y_1,\ldots,y_n$ be real-analytic delay functions near zero.
Given their local identity classes and the vanishing orders of the inter-class delay differences, $H_\lambda$
equals the optimal value of an integer minimum-cost flow with $O(n^2)$ nodes and arcs
(taken as $+\infty$ if infeasible).
Therefore, in polynomial time in $n$ and in the binary bit lengths of the given orders,
one can determine whether $p_\lambda$ vanishes identically and, if not, obtain the exact vanishing order $2H_\lambda$.
\end{theorem}

The paper is organized as follows. Section~\ref{sec2} introduces the notation and preliminary lemmas. In Section~\ref{sec3}, we prove Theorems~\ref{thm1}--\ref{thm4}.
Section~\ref{sec4} provides further discussion.

\section{Preliminaries}\label{sec2}

Let $\mathbb{N}_0=\{0,1,2,\ldots\}$, and let $I$ be the identity operator.
Let $\mathcal{H}_{\rm ext}=\mathbb{C}^M$ be the Hilbert space for the spatial degrees of freedom of a single photon, where $M\geq n$ is the number of spatial channels of the optical network.
Channels in this paper always mean spatial modes of the optical network, and the orthonormal basis $|1\rangle,\ldots,|M\rangle$ corresponds to these spatial modes, respectively.
We fix the photon polarization and describe its internal state by frequency. Let
$\mathcal{H}_{\rm int}=L^2(\mathbb{R},d\nu)$ be the internal-state Hilbert space,
where $\nu$ denotes the angular-frequency offset relative to a common central angular frequency.
A single-photon state belongs to $\mathcal{H}_{\rm ext}\otimes\mathcal{H}_{\rm int}$.
We adopt the convention that the complex inner product is conjugate-linear in the first argument and linear in the second.
In $\mathcal{H}_{\rm int}$,
$\langle f,g\rangle=\int_\mathbb{R}\overline{f(\nu)}g(\nu)\,d\nu$,
and the corresponding norm satisfies $\|v\|^2=\langle v,v\rangle$.
Denote by $|v\rangle\langle w|$ the operator $u\mapsto v\langle w,u\rangle$.
The inner product on tensor products satisfies
\[
 \left\langle\bigotimes_{i=1}^n u_i,\bigotimes_{i=1}^n v_i\right\rangle
 =\prod_{i=1}^n\langle u_i,v_i\rangle.
\]
The $n$-fold tensor product of the same space $\mathcal{H}$ is denoted by $\mathcal{H}^{\otimes n}$.
The positions of tensor factors are used only to represent the joint quantum state and do not assign observable individual labels to identical photons.
Let $A^*$ and $\operatorname{tr} A$ denote the adjoint and trace of an operator $A$, respectively.
If $A=A^*$ and $\langle v,Av\rangle\geq0$ for all $v$, then $A$ is called positive semidefinite; $A\leq B$ means that $B-A$ is positive semidefinite.
A unitary operator satisfies $U^*U=UU^*=I$, and an orthogonal projection satisfies $P=P^*=P^2$.
We now introduce the photon input and measurement model.
Let $n$ independent single photons enter through mutually distinct channels $i_1,\ldots,i_n$, with identical internal states except for time delays $\theta_i$.
In the idealized model of this paper, the network is a lossless passive linear optical network described, for all frequencies, by the same unitary matrix $U$;
$U_{ab}$ is the transition amplitude from input channel $b$ to output channel $a$, and the multiphoton spatial transformation is $U^{\otimes n}$.
The detectors perform photon-number measurements over the entire wave packet and do not resolve the internal degrees of freedom.
Let $s>0$ be the angular-frequency standard deviation of the common Gaussian spectral probability density.
The corresponding spectral probability density and the internal-state wave function of the $i$-th photon are respectively
\[
 q_s(\nu)=\frac{e^{-\nu^2/(2s^2)}}{\sqrt{2\pi}s},\qquad
 \phi_i(\nu)=\sqrt{q_s(\nu)}e^{-\mathrm i\theta_i\nu}.
\]
The above wave functions satisfy $\|\phi_i\|=1$. Let $y_i=s\theta_i$ be the dimensionless delay,
and form the Gram matrix from the pairwise inner products of the internal states:
\[
 G(\boldsymbol{y})_{ij}=\langle\phi_i,\phi_j\rangle
 =\int_{\mathbb{R}}q_s(\nu)e^{\mathrm i(\theta_i-\theta_j)\nu}\,d\nu
 =e^{-(y_i-y_j)^2/2}.
\]
$G$ is positive semidefinite, its diagonal entries are $1$, and its off-diagonal entries characterize the internal-state overlaps of the corresponding photons.
Adding the same shift to all delays leaves $G$ unchanged. If the common central angular frequency $\omega_0$ is retained,
then the matrix elements are merely multiplied by the phase factor
$e^{\mathrm i\omega_0(\theta_i-\theta_j)}$, which cancels in each permutation product and hence does not change the immanant.
The dimensionless time-mismatch parameter in Eq.~\eqref{eq2} satisfies
\[
 S=s^2\sum_i(\theta_i-\bar\theta)^2
   =\frac1n\sum_{i<j}(y_i-y_j)^2,
 \qquad \bar\theta=\frac1n\sum_i\theta_i.
\]
The Gaussian condition here describes only the spectral shape of a single photon; each occupied input channel still contains exactly one photon.
A real-analytic delay function means a delay function that can be represented near the origin by a convergent Taylor series.
Two delay functions are locally identical if they agree in some neighborhood of the origin.
A positive semidefinite operator with trace $1$ is called a density operator and is used to describe a photon state.
A unit vector $\psi$ corresponds to the pure state
$|\psi\rangle\langle\psi|$.
Taking the partial trace of a joint state $\rho$ over the internal degrees of freedom gives the spatial reduced state
$\rho_{\rm ext}=\operatorname{tr}_{\rm int}\rho$. This operator is uniquely determined by the condition
\[
 \operatorname{tr}(A\rho_{\rm ext})=\operatorname{tr}\bigl((A\otimes I)\rho\bigr)
\]
for all spatial operators $A$, and therefore it preserves all measurement probabilities involving only spatial degrees of freedom.
All spatial operators in this paper are finite-dimensional operators. The support of a positive semidefinite operator is the sum of the eigenspaces corresponding to its nonzero eigenvalues.
The trace norm of an operator $A$ is defined as $\|A\|_1=\operatorname{tr}\sqrt{A^*A}$;
if $A$ is self-adjoint, its trace norm equals the sum of the absolute values of its eigenvalues.
The trace distance between two density operators $\rho$ and $\sigma$ is $\|\rho-\sigma\|_1/2$.
A complete output photon-number configuration is denoted by $o=(m_1,\ldots,m_M)$, where $m_a\in\mathbb{N}_0$ is the photon number in the $a$-th channel and $\sum_a m_a=n$.
The measurement projection corresponding to this outcome is
\[
 M_o=\sum_{\substack{j_1,\ldots,j_n\in\{1,\ldots,M\}\\
           \#\{r:j_r=a\}=m_a\ (1\leq a\leq M)}}
 |j_1,\ldots,j_n\rangle\langle j_1,\ldots,j_n|.
\]
These operators satisfy $M_o\geq0$, $\sum_oM_o=I$,
and the counting probability is $\operatorname{tr}(M_oU^{\otimes n}\rho_{\rm ext}(U^{\otimes n})^*)$.
The above counting outcomes include multiphoton events with $m_a\geq2$ in the same output channel.
To describe exchange symmetry, consider the action of $S_n$ on $\mathcal{H}^{\otimes n}$ by permuting tensor factors:
\[
 R(\sigma)(v_1\otimes\cdots\otimes v_n)
 =v_{\sigma^{-1}(1)}\otimes\cdots\otimes v_{\sigma^{-1}(n)}.
\]
This action gives a representation of $S_n$. A nonzero representation is called irreducible if it has no nonzero proper invariant subspace.
The complex irreducible representations of $S_n$ are labeled by partitions $\lambda\vdash n$,
the character $\chi_{\lambda}(\sigma)$ is the trace of the corresponding representation matrix,
and $d_\lambda=\chi_{\lambda}(e)$ is the dimension of the representation.
By exchange-symmetry type in this paper we mean the irreducible representation type under the tensor-factor permutation action.
The sum of all irreducible subspaces isomorphic to type $\lambda$ is called the $\lambda$-isotypic subspace,
and its orthogonal projection is
\[
 E_\lambda=\frac{d_\lambda}{n!}
             \sum_{\sigma\in S_n}\chi_{\lambda}(\sigma)R(\sigma).
\]
Characters of the symmetric group are real-valued and invariant under taking inverses.
By the character orthogonality relations,
$E_\lambda^*=E_\lambda$, $E_\lambda^2=E_\lambda$,
$E_\lambda E_\mu=0$ ($\lambda\ne\mu$), $\sum_\lambda E_\lambda=I$,
see \cite{sagan}. The subscripts or superscripts $\rm ext$ and $\rm int$ indicate that the corresponding operator acts on tensor powers of the spatial and internal spaces, respectively.
The overall quantum state of the photons satisfies Bose exchange symmetry
\[
 (R_{\rm ext}(\sigma)\otimes R_{\rm int}(\sigma))\Psi=\Psi
 \qquad(\sigma\in S_n),
\]
and after taking the partial trace over the internal degrees of freedom, the spatial reduced state can contain different exchange-symmetry types.

\begin{lemma}\label{lem1}
Let $\phi_1,\ldots,\phi_n$ be normalized internal-state vectors, $\Phi=\bigotimes_i\phi_i$,
$G_{ij}=\langle\phi_i,\phi_j\rangle$.
If the photons enter through distinct spatial channels, then their spatial reduced state $\rho_{\rm ext}$ satisfies
\[
 p_\lambda:=\frac{d_\lambda}{n!}\operatorname{Imm}_\lambda(G)
 =\|E_\lambda\Phi\|^2
 =\operatorname{tr}(E_\lambda^{\rm ext}\rho_{\rm ext}),\qquad
 p_\lambda\geq0,\quad\sum_{\lambda\vdash n}p_\lambda=1.
\]
Moreover, $\rho_{\rm ext}$ commutes with all $E_\lambda^{\rm ext}$.
The network transformation $U^{\otimes n}$, which is independent of the internal states, preserves the total weight $p_\lambda$ of each exchange-symmetry type.
\end{lemma}
\begin{proof}
From the properties of orthogonal projections and the definition of the central projection, we obtain
\[
 \|E_\lambda\Phi\|^2
 =\frac{d_\lambda}{n!}\sum_\sigma\chi_{\lambda}(\sigma)
                 \prod_iG_{i,\sigma^{-1}(i)}.
\]
Replacing $\sigma$ by $\sigma^{-1}$ yields the first equality.
Nonnegativity and the fact that the sum equals $1$ then follow.
Let $|\boldsymbol{i}\rangle=|i_1\rangle\otimes\cdots\otimes|i_n\rangle$.
The overall bosonic symmetric state is
\[
 |\Psi\rangle=\frac1{\sqrt{n!}}\sum_{\sigma\in S_n}
 R_{\rm ext}(\sigma)|\boldsymbol{i}\rangle\otimes R_{\rm int}(\sigma)\Phi,
 \qquad \rho_{\rm ext}=\operatorname{tr}_{\rm int}|\Psi\rangle\langle\Psi|.
\]
Since the input spatial channels are distinct, the spatial state vectors $R_{\rm ext}(\sigma)|\boldsymbol{i}\rangle$ are pairwise orthogonal,
so $|\Psi\rangle$ is normalized.
The overall state is invariant under simultaneous permutations of the spatial and internal tensor factors, that is,
$(R_{\rm ext}(g)\otimes R_{\rm int}(g))|\Psi\rangle=|\Psi\rangle$.
The partial trace is invariant under unitary conjugation on the internal space, so
\[
 R_{\rm ext}(g)\rho_{\rm ext}R_{\rm ext}(g)^*=\rho_{\rm ext}.
\]
This proves that $\rho_{\rm ext}$ commutes with the spatial tensor-factor permutations and hence with their linear combinations $E_\lambda^{\rm ext}$.
For any $g\in S_n$, reindexing the sum in the overall bosonic state gives
\[
 (R_{\rm ext}(g)\otimes I)|\Psi\rangle
 =(I\otimes R_{\rm int}(g^{-1}))|\Psi\rangle.
\]
Averaging over $g$ with the character gives
$(E_\lambda^{\rm ext}\otimes I)|\Psi\rangle
=(I\otimes E_\lambda^{\rm int})|\Psi\rangle$.
Using the orthogonality of the spatial state vectors, the squared norm of the vector on the right-hand side is
$\frac1{n!}\sum_\sigma\|E_\lambda R_{\rm int}(\sigma)\Phi\|^2
=\|E_\lambda\Phi\|^2$.
Finally, $U^{\otimes n}$ commutes with the spatial tensor-factor permutations and their central projections, and therefore preserves the weights $p_\lambda$ of the exchange-symmetry components.
\end{proof}

Lemma~\ref{lem1} shows that $p_\lambda$ is the weight of the corresponding exchange-symmetry component in the spatial reduced state.
When $p_\lambda>0$, let
\[
 \rho_\lambda=
 \frac{E_\lambda^{\rm ext}\rho_{\rm ext}E_\lambda^{\rm ext}}{p_\lambda}.
\]
Using the commutation property, we have $\rho_{\rm ext}=\sum_{\lambda:p_\lambda>0}p_\lambda\rho_\lambda$.
Let $V=U^{\otimes n}$; then the probability of any output event is
\[
 P(o)=\sum_{\lambda:p_\lambda>0}
       p_\lambda\,\operatorname{tr}(M_oV\rho_\lambda V^*).
\]
Thus, the weights of the exchange-symmetry components are determined by the internal-state overlaps of the input photons, while the probabilities of specific output counting events also depend on the normalized density operators of the components and on the optical network.
This decomposition does not alter the Bose statistics of the photons.
The product vector $\Phi$ in the lemma is used only to compute the weights; the complete physical state must still be globally symmetrized.
In particular, for the mutually distinct input channels above, let
\[
 |\psi_{\rm sym}\rangle
 =\frac1{\sqrt{n!}}\sum_{\sigma\in S_n}
       R_{\rm ext}(\sigma)|\boldsymbol{i}\rangle.
\]
The orthogonality of the spatial state vectors ensures that $|\psi_{\rm sym}\rangle$ is normalized.
Since
$E_{(n)}^{\rm ext}R_{\rm ext}(\sigma)|\boldsymbol{i}\rangle
=|\psi_{\rm sym}\rangle/\sqrt{n!}$,
projecting the sum in the overall state term by term gives
\[
 (E_{(n)}^{\rm ext}\otimes I)|\Psi\rangle
 =|\psi_{\rm sym}\rangle\otimes E_{(n)}^{\rm int}\Phi.
\]
Taking the partial trace over the internal degrees of freedom and then using Lemma~\ref{lem1}, we obtain
\[
 \begin{aligned}
 E_{(n)}^{\rm ext}\rho_{\rm ext}E_{(n)}^{\rm ext}
 &=\|E_{(n)}^{\rm int}\Phi\|^2
       |\psi_{\rm sym}\rangle\langle\psi_{\rm sym}|\\
 &=p_{(n)}|\psi_{\rm sym}\rangle\langle\psi_{\rm sym}|.
 \end{aligned}
\]
In the Gaussian model of this paper, all entries of $G$ are positive,
so $p_{(n)}=\operatorname{per}(G)/n!>0$.
Therefore, after retaining only the fully symmetric component and normalizing it,
$\rho_{(n)}=|\psi_{\rm sym}\rangle\langle\psi_{\rm sym}|$,
independently of the time delays.
This spatial state is exactly the spatial state of ideally indistinguishable photons under the same input-channel configuration,
so the two have the same complete output photon-number distribution after passing through the same network.
Two-photon interference concretely illustrates the distinct roles of the exchange-symmetry component weights and the network transition amplitudes.
Let $A$ be the $2\times2$ transition submatrix corresponding to two distinct input channels and two distinct output channels,
and let $\gamma=|G_{12}|^2$. Then the probability that the two photons appear in these two output channels, respectively, is
\[
 P(\boldsymbol{o})=\frac{1+\gamma}{2}|\operatorname{per}A|^2
          +\frac{1-\gamma}{2}|\det A|^2.
\]
The two coefficients are $p_{(2)}$ and $p_{(1,1)}$, respectively.
For a balanced beam splitter $A=2^{-1/2}\left(\begin{smallmatrix}1&1\\1&-1\end{smallmatrix}\right)$,
we have $\operatorname{per}A=0$ and $|\det A|=1$, so the HOM coincidence probability is
$P(1,1)=(1-\gamma)/2$, see \cite{till}.
For multiphoton systems, exchange-symmetry types of dimension greater than $1$ also involve matrix elements within the corresponding isotypic subspaces,
and the formula cannot be directly generalized to a sum of products of the individual $p_\lambda$ and the squared modulus of a single immanant,
see \cite{till} and \cite{khalid}.
Following the notation for Young diagrams and tableaux in the introduction, the conjugate partition satisfies $\lambda'_j=\#\{r:\lambda_r\geq j\}$.
By the Vandermonde determinant formula
\[
 \det[z_u^{r-1}]_{r,u=1}^{h}=\prod_{1\leq r<u\leq h}(z_u-z_r),
\]
taking determinants column by column and multiplying them gives $\Delta_T$.
Here $b_\lambda$ is the total number of pairs of distinct boxes within the same column,
$c_\lambda$ is the product of the numbers of permutations within the columns, and $q_\lambda$ is the factorial factor appearing in the lowest-degree expansion.

Two tableaux are called row-equivalent if each row contains the same set of labels.
The row-equivalence class of a tableau $T$ is denoted by $[T]$ and is called a row tabloid.
The vector space with these row tabloids as an orthonormal basis is denoted by $M^\lambda$,
and $S_n$ acts on this space by permuting labels.
Let $C_j(T)$ be the set of labels in the $j$-th column of $T$,
and let $C_T$ be the group of label permutations that preserve each column set.
As $T$ runs over $\mathcal{T}_\lambda$, the column-alternating vectors
$e_T=\sum_{\pi\in C_T}\operatorname{sgn}(\pi)[\pi T]$
span the irreducible Specht module $S^\lambda$ of dimension $d_\lambda$.
Here a module means a vector space equipped with a linear group action.
Schur's lemma states that an operator on a complex irreducible module that commutes with the group action must be a scalar multiple of the identity operator.

We use nonnegative integers as entries of semistandard tableaux, requiring rows to be weakly increasing and columns to be strictly increasing.
The content of a semistandard tableau is denoted by $m=(m_0,m_1,\ldots)$, where $m_j$ records the number of occurrences of the entry $j$,
satisfying $m_j\in\mathbb{N}_0$ and $\sum_jm_j=n$.
Taking all nonnegative integer sequences of length $n$ with content $m$ as a basis,
and letting $S_n$ permute the positions of the sequences, gives the permutation module $M^m$.
Young's rule gives its irreducible decomposition
\[
 M^m\cong\bigoplus_{\rho\vdash n}K_{\rho m}S^\rho,
 \qquad K_{\rho m}>0\ \Longleftrightarrow\ \rho\succeq m^+,
 \qquad K_{\lambda\lambda}=1.
\]
Here $m^+$ is the partition obtained by arranging the nonzero multiplicities in $m$ in nonincreasing order,
and $K_{\rho m}$ is the number of semistandard tableaux of shape $\rho$ and content $m$,
which also equals the multiplicity of $S^\rho$ in $M^m$;
the summand $K_{\rho m}S^\rho$ in the direct sum denotes the corresponding number of isomorphic copies.
In particular, the type-$\lambda$ copy in $M^\lambda$ is unique.
The irreducibility of Specht modules, Schur's lemma, and Young's rule can all be found in \cite{sagan}.

Finally, for a multi-index $\alpha=(\alpha_1,\ldots,\alpha_n)\in\mathbb{N}_0^n$, write
\[
 |\alpha|=\sum_i\alpha_i,\qquad
 \alpha!=\prod_i\alpha_i!,\qquad
 \boldsymbol{x}^\alpha=\prod_i x_i^{\alpha_i}.
\]
In the square-summable sequence space $\ell^2(\mathbb{N}_0)$ used later,
take the orthonormal basis $\{e_k:k\in\mathbb{N}_0\}$,
and let $e_\alpha=e_{\alpha_1}\otimes\cdots\otimes e_{\alpha_n}$.
Here $|\alpha|$ denotes the total degree of the expansion, not the photon number; the photon number is always $n$.
The following lemma determines the minimal degree of occurrence of each type and the corresponding projection constant.

\begin{lemma}\label{lem2}
Let $\lambda\vdash n$, $\boldsymbol{x}\in\mathbb{R}^n$, let $\{e_k:k\geq0\}$ be an orthonormal sequence of vectors,
and let $e_\alpha=e_{\alpha_1}\otimes\cdots\otimes e_{\alpha_n}$ and $|\alpha|=\sum_i\alpha_i$.
If $E_\lambda e_\alpha\ne0$, then $|\alpha|\geq b_\lambda$.
When equality holds, necessarily $\#\{i:\alpha_i=j\}=\lambda_{j+1}$ for all $j\geq0$.
Let
\[
 w_\lambda(\boldsymbol{x})=
 \sum_{\alpha:\,\#\{i:\alpha_i=j\}=\lambda_{j+1}\ (\forall j)}
                  \boldsymbol{x}^\alpha e_\alpha,
 \qquad \boldsymbol{x}^\alpha=\prod_i x_i^{\alpha_i}.
\]
Then
\[
 \|E_\lambda w_\lambda(\boldsymbol{x})\|^2
 =\frac{d_\lambda}{n!c_\lambda}
                \sum_{T\in\mathcal{T}_\lambda}\Delta_T(\boldsymbol{x})^2.
\]
\end{lemma}
\begin{proof}
Let $m_j=\#\{i:\alpha_i=j\}$.
The permutation orbit of the basis tensor $e_\alpha$ spans $M^m$.
If its type-$\lambda$ projection is nonzero, Young's rule guarantees the existence of a semistandard tableau of shape $\lambda$ and content $m$.
Every entry in the $r$-th row of this tableau is at least $r-1$, because above it in the same column there are $r-1$ strictly smaller nonnegative integers.
Hence the sum of entries $|\alpha|$ is at least $b_\lambda$.
Equality requires that the $r$-th row be filled entirely with $r-1$, yielding the unique content stated above.
This means that at total degree $b_\lambda$, all content spaces except this unique content are annihilated by $E_\lambda$.
Within the content space corresponding to the above minimal total degree, mapping $[T]$ to the basis tensor satisfying
$\alpha_i=r-1$ when label $i$ lies in row $r$ gives an isometric isomorphism with $M^\lambda$.
The row tabloids $[\pi T]$ produced by column permutations are pairwise distinct.
Indeed, a label permutation that simultaneously preserves every row set and every column set can only fix every box, and hence must be the identity permutation.
The column group has $c_\lambda$ elements, so $\|e_T\|^2=c_\lambda$.
Let $e_Te_T^*$ denote the rank-one operator $v\mapsto e_T\langle e_T,v\rangle$.
The operator $F=\sum_{T\in\mathcal{T}_\lambda}e_Te_T^*$ commutes with label permutations,
and its image is contained in the unique $S^\lambda$ submodule.
Since $F$ is self-adjoint, it vanishes on the orthogonal complement of this submodule;
Schur's lemma further implies that on the irreducible submodule $S^\lambda$ it must be a scalar multiple of the identity.
The trace of each rank-one term is $\|e_T\|^2$, and the sum runs over $n!$ tableaux, so $\operatorname{tr} F=n!c_\lambda$.
Together with $\operatorname{tr} E_\lambda|_{M^\lambda}=d_\lambda$, this gives
\[
 F=\frac{n!c_\lambda}{d_\lambda}E_\lambda\big|_{M^\lambda}.
\]
Alternating column by column and using the Vandermonde determinant formula gives
\[
 \langle e_T,w_\lambda(\boldsymbol{x})\rangle
 =\prod_j\det[x_{T(u,j)}^{r-1}]_{r,u=1}^{\lambda'_j}
 =\Delta_T(\boldsymbol{x}).
\]
Substituting $w_\lambda$ into the quadratic form of $F$ gives
\[
 \|E_\lambda w_\lambda(\boldsymbol{x})\|^2
 =\frac{d_\lambda}{n!c_\lambda}
                \sum_{T\in\mathcal{T}_\lambda}\Delta_T(\boldsymbol{x})^2.
\]
This completes the proof.
\end{proof}

\section{Proofs of the Theorems}\label{sec3}

\begin{proof}[\normalfont\bfseries Proof of Theorem~\ref{thm1}]
Replacing each $x_i$ by $x_i-\bar x$ does not change the Gram matrix or the column-difference product.
Therefore, without loss of generality, let $\sum_i x_i=0$,
$S=\sum_i x_i^2$, and $t=\varepsilon^2$.
Take an orthonormal basis $e_0,e_1,\ldots$ in $\ell^2(\mathbb{N}_0)$ and let
\[
 v_i(t)=e^{-t x_i^2/2}\sum_{k\geq0}\frac{t^{k/2}x_i^k}{\sqrt{k!}}e_k.
\]
The series converges in the Hilbert-space norm, and
$\langle v_i(t),v_j(t)\rangle=e^{-t(x_i-x_j)^2/2}$.
By Lemma~\ref{lem1}, the weights are determined solely by the Gram matrix, so these vectors may be used in place of the physical internal states.
Here $e_k$ is a basis of an auxiliary internal-state space. Since $t=\varepsilon^2\geq0$,
$t^{1/2}$ is taken as the nonnegative square root; the resulting Gram matrix is independent of the sign of $\varepsilon$.
For $\alpha!=\prod_i\alpha_i!$, let
$f_m(\boldsymbol{x})=\sum_{|\alpha|=m}\boldsymbol{x}^\alpha e_\alpha/\sqrt{\alpha!}$.
Decompose the tensor space according to the total degree $\sum_i\alpha_i$. The components are mutually orthogonal and invariant under permutations; hence
\begin{equation}
 \begin{aligned}
 p_\lambda(\varepsilon\boldsymbol{x})
   &=e^{-tS}\sum_{m\geq0}a_{\lambda,m}(\boldsymbol{x})t^m,\\
 a_{\lambda,m}(\boldsymbol{x})&=\|E_\lambda f_m(\boldsymbol{x})\|^2\geq0,\\
 \sum_{\lambda\vdash n}a_{\lambda,m}(\boldsymbol{x})&=\|f_m(\boldsymbol{x})\|^2=\frac{S^m}{m!}.
 \end{aligned}
 \label{eq7}
\end{equation}
Here the total degree of the expansion does not represent the photon number; the photon number remains $n$.
There are no cross terms between different degrees in Eq.~\eqref{eq7}; the last equation follows from the completeness of the projections and the polynomial expansion identity
\[
 \sum_{|\alpha|=m}\frac{\boldsymbol{x}^{2\alpha}}{\alpha!}
 =\frac{(x_1^2+\cdots+x_n^2)^m}{m!}.
\]
Moreover, $\sum_m t^mS^m/m!=e^{tS}$ implies that the series converges absolutely,
so the interchange of summation and projection above is valid.
By Lemma~\ref{lem2}, $a_{\lambda,m}=0$ when $m<b_\lambda$;
when $m=b_\lambda$, only the component with content $\lambda$ contributes, and $\alpha!$ is identically $q_\lambda$.
The sum-of-squares formula in that lemma then gives $a_{\lambda,b_\lambda}=\kappa_\lambda$.
Since the remaining coefficients are nonnegative,
\[
 p_\lambda(\varepsilon\boldsymbol{x})
 \geq e^{-tS}\kappa_\lambda(\boldsymbol{x})t^{b_\lambda}.
\]
Taking $\varepsilon=1$ and choosing $\boldsymbol{x}$ to be the centered $\boldsymbol{y}$ gives the lower bound in Eq.~\eqref{eq3}.
Also, by Eq.~\eqref{eq7}, for fixed $\boldsymbol{x}$ and $t\to0^+$,
\[
 0\leq\sum_{m\geq b_\lambda+1}a_{\lambda,m}(\boldsymbol{x})t^m
 \leq\sum_{m\geq b_\lambda+1}\frac{(tS)^m}{m!}
 =O(t^{b_\lambda+1}).
\]
Combining this with $e^{-tS}=1+O(t)$ and $t=\varepsilon^2$ yields the stated asymptotic expansion.

We now prove the upper bound. In the tensor-permutation representation, let
\[
 Q_T=\frac1{c_\lambda}\sum_{\sigma\in C_T}\operatorname{sgn}(\sigma)R(\sigma),
 \qquad B_\lambda=\frac1{n!}\sum_{T\in\mathcal{T}_\lambda}Q_T.
\]
Each $Q_T$ is the orthogonal projection onto the sign-isotypic subspace for the column group $C_T$, and
$R(g)Q_TR(g)^{-1}=Q_{gT}$.
Therefore $B_\lambda$ corresponds to a central element of the group algebra and acts as the same scalar on all copies of irreducible $S^\mu$ of the same type.
This scalar multiplied by $d_\mu$ equals the trace of $Q_T$ on $S^\mu$, i.e., the multiplicity of the sign representation in the restriction of $S^\mu$ to $C_T$.
By Frobenius reciprocity and Young's rule, this multiplicity equals the multiplicity $K_{\mu',\lambda'}$ of $S^\mu$ in
\[
 \operatorname{Ind}_{C_T}^{S_n}\operatorname{sgn}
 \cong\operatorname{sgn}\otimes M^{\lambda'}
 \cong\bigoplus_{\rho\vdash n}K_{\rho,\lambda'}S^{\rho'},
\]
where we used the sign-twist relation $\operatorname{sgn}\otimes S^\rho\cong S^{\rho'}$, see \cite{sagan}.
Hence
\[
 B_\lambda=\sum_{\mu\vdash n}\frac{K_{\mu',\lambda'}}{d_\mu}E_\mu,
 \qquad E_\lambda\leq d_\lambda B_\lambda,
\]
where the second inequality holds because $K_{\lambda',\lambda'}=1$
and the remaining multiplicities are nonnegative, so $d_\lambda B_\lambda-E_\lambda$ is positive semidefinite.

Let $\Phi=\bigotimes_i v_i(1)$ and choose $\boldsymbol{x}$ to be the centered $\boldsymbol{y}$.
Taking the expectation of the above operator inequality in the vector $\Phi$ and expanding the determinants corresponding to the columns gives
\[
 p_\lambda(\boldsymbol{y})
 \leq\frac{d_\lambda}{n!c_\lambda}
       \sum_{T\in\mathcal{T}_\lambda}\prod_j\det G[C_j(T)].
\]
Here $G[C]$ denotes the principal submatrix with row and column indices in $C$.
Indeed, the column group is the direct product of the symmetric groups of the columns, so
\[
 \langle\Phi,Q_T\Phi\rangle
 =\frac1{c_\lambda}\prod_j
   \left(\sum_{\sigma_j\in S_{C_j(T)}}\operatorname{sgn}(\sigma_j)
          \prod_{i\in C_j(T)}G_{i,\sigma_j(i)}\right)
 =\frac1{c_\lambda}\prod_j\det G[C_j(T)].
\]
For $h$ real numbers $z_1,\ldots,z_h$, let $Z=\operatorname{diag}(z_1,\ldots,z_h)$ and
$Q_h=\prod_{k=0}^{h-1}k!$.
Let $dU$ be the Haar probability measure on the unitary group $\mathrm U(h)=\{U:U^*U=I\}$;
it has total mass $1$ and is invariant under left and right multiplication by fixed unitary matrices. By the Harish--Chandra--Itzykson--Zuber (HCIZ) integral formula (see \cite{iz,mcswiggen}),
\[
 \det[e^{z_i z_j}]
 =\frac{\prod_{i<j}(z_j-z_i)^2}{Q_h}
       \int_{\mathrm U(h)}e^{\operatorname{tr}(ZUZU^*)}\,dU.
\]
Factoring $e^{-z_i^2/2}$ and $e^{-z_j^2/2}$ out of the $i$-th row and $j$-th column of the Gaussian kernel matrix, respectively, gives
\[
 \det[e^{-(z_i-z_j)^2/2}]
 =\frac{\prod_{i<j}(z_j-z_i)^2}{Q_h}
   e^{-\operatorname{tr} Z^2}\int_{\mathrm U(h)}e^{\operatorname{tr}(ZUZU^*)}\,dU
 \leq\frac{\prod_{i<j}(z_j-z_i)^2}{Q_h}.
\]
Here, by the Cauchy--Schwarz inequality for the Hilbert--Schmidt inner product $\langle A,B\rangle_{\rm HS}=\operatorname{tr}(A^*B)$, we have
\[
 \operatorname{tr}(ZUZU^*)\leq
 \sqrt{\operatorname{tr} Z^2}\sqrt{\operatorname{tr}\bigl((UZU^*)^2\bigr)}=\operatorname{tr} Z^2.
\]
Therefore, after multiplying the integrand by $e^{-\operatorname{tr} Z^2}$, it is at most $1$, and the required inequality follows immediately.
The above HCIZ equality holds for distinct coordinates and extends by continuity to the case of coincident coordinates.
Substituting this estimate into the upper bound for the principal minors and using
$\prod_j Q_{\lambda'_j}=q_\lambda$ gives $p_\lambda\leq\kappa_\lambda$.

If $\lambda\succeq\nu$, Young's rule implies the existence of a semistandard tableau of shape $\lambda$ and content $\nu$.
Group photons with equal delays, number the groups according to their sizes, and then replace the corresponding entries by distinct photon labels within each group.
Since each column of a semistandard tableau is strictly increasing, the coordinates in each column of the resulting tableau $T$ are distinct.
Hence $\Delta_T(\boldsymbol{x})\ne0$, and therefore $\kappa_\lambda(\boldsymbol{x})>0$.
If $\lambda\not\succeq\nu$, then for any real $\varepsilon$, let
$\Phi_\varepsilon=\bigotimes_i v_i(\varepsilon^2)$.
This tensor is invariant under the subgroup
$H\cong S_{\nu_1}\times\cdots\times S_{\nu_{\ell(\nu)}}$ that permutes labels with equal delays.
Let $\mathcal{V}_\varepsilon$ be the submodule spanned by its permutation orbit; then the map
\[
 \mathbb{C}[S_n/H]\longrightarrow\mathcal{V}_\varepsilon,\qquad
 [\sigma H]\longmapsto R(\sigma)\Phi_\varepsilon
\]
is a surjective module homomorphism.
Indeed, the fact that $H$ fixes $\Phi_\varepsilon$ ensures that the map is well defined; it is compatible with the left group action, and its image contains all orbit vectors spanning $\mathcal{V}_\varepsilon$.
Therefore $\mathcal{V}_\varepsilon$ is a quotient module of $\mathbb{C}[S_n/H]\cong M^\nu$.
By Young's rule, $M^\nu$ contains no $S^\lambda$, so its quotient modules also contain no copy of this type.
Thus $E_\lambda\Phi_\varepsilon=0$,
i.e., $p_\lambda(\varepsilon\boldsymbol{x})=0$ for all $\varepsilon$.

Finally, substituting $\boldsymbol{y}=\varepsilon\boldsymbol{x}$ into Eq.~\eqref{eq3} and using the homogeneity of $\kappa_\lambda$ of degree $2b_\lambda$ and $1-e^{-u}\leq u$ ($u\geq0$) yields the uniform error bound in the theorem.
\end{proof}

\begin{proof}[\normalfont\bfseries Proof of Theorem~\ref{thm2}]
For tableaux $T$ with $w(T)<\infty$, multiplying the leading-order expansions of the delay differences gives
$\Delta_T(\boldsymbol{y}(\varepsilon))
=A_T\varepsilon^{w(T)}+O(|\varepsilon|^{w(T)+1})$;
if $w(T)=+\infty$, then the column-difference product is identically zero.
When $H_\lambda<\infty$, substituting these expansions into Eq.~\eqref{eq2} gives
\[
 \kappa_\lambda(\boldsymbol{y}(\varepsilon))
 =K_\lambda\varepsilon^{2H_\lambda}
           +O(|\varepsilon|^{2H_\lambda+1}).
\]
At least one tableau satisfies $w(T)=H_\lambda$, and its coefficient $A_T$ is nonzero. Hence the leading coefficient $K_\lambda$ of the sum of squares is positive.
By the continuity of $S(\boldsymbol{y}(\varepsilon))$ and Eq.~\eqref{eq3}, there exist constants $c,C>0$ such that
\[
 c|\varepsilon|^{2H_\lambda}\leq p_\lambda(\varepsilon)
          \leq C|\varepsilon|^{2H_\lambda}
\]
holds for sufficiently small nonzero $\varepsilon$.
Moreover, $p_\lambda$ is a linear combination of finitely many analytic exponential functions and hence is real analytic.
The above lower bound shows that it is not identically zero. Let $r_0=\operatorname{ord}_0p_\lambda$; then
\[
 p_\lambda(\varepsilon)
 =c_{r_0}\varepsilon^{r_0}+O(|\varepsilon|^{r_0+1}),
 \qquad c_{r_0}\ne0.
\]
Take $\varepsilon>0$ and divide by $\varepsilon^{2H_\lambda}$.
If $r_0<2H_\lambda$, the above upper bound cannot hold;
if $r_0>2H_\lambda$, the above positive lower bound cannot hold.
Hence $r_0=2H_\lambda$ and $c_{r_0}>0$.
Letting $C_\lambda=c_{r_0}$ gives the stated expansion.
Dividing Eq.~\eqref{eq3} by $\varepsilon^{2H_\lambda}$ and letting $\varepsilon\to0$ gives the two-sided bound for $C_\lambda$.
At the completely synchronized point, $S(\boldsymbol{y}(0))=0$, so $C_\lambda=K_\lambda$.

If $H_\lambda=+\infty$, then all column-difference products are identically zero, and Eq.~\eqref{eq3} gives $p_\lambda\equiv0$.
For two delay functions that are not locally identical, their difference is nonzero in a sufficiently small punctured neighborhood.
Since the number of photon pairs is finite, one can choose a common neighborhood such that, at every nonzero $\varepsilon$, the multiplicities of equal delays are all $\nu$.
Applying the nonzero criterion of Theorem~\ref{thm1} then gives the final equivalence.
\end{proof}

\begin{proof}[\normalfont\bfseries Proof of Theorem~\ref{thm3}]
Center $\boldsymbol{y}$ and take $\boldsymbol{x}=\boldsymbol{y}$, $t=1$ in Eq.~\eqref{eq7}.
Since $a_{\lambda,m}=0$ for $m<b_\lambda$, interchanging the order of summation in the nonnegative series gives
\[
 \delta_K=e^{-S}\sum_{m\geq K}\sum_{\lambda:\,b_\lambda\geq K}a_{\lambda,m}
 \leq e^{-S}\sum_{m\geq K}\frac{S^m}{m!}.
\]
Also, $(K+r)!\geq K!r!$, so
$e^{-S}\sum_{m\geq K}S^m/m!\leq(S^K/K!)e^{-S}\sum_{r\geq0}S^r/r!=S^K/K!$,
which proves Eq.~\eqref{eq5}.

Since $b_{(n)}=0<K$, the fully symmetric type is included in the retained part.
For $\lambda=(n)$, we have $d_\lambda=c_\lambda=q_\lambda=1$,
the column-difference product of every tableau is $1$, and $|\mathcal{T}_{(n)}|=n!$.
Therefore Eq.~\eqref{eq2} gives $\kappa_{(n)}=n!/n!=1$.
By Theorem~\ref{thm1}, $p_{(n)}\geq e^{-S}\kappa_{(n)}=e^{-S}>0$,
so $1-\delta_K\geq e^{-S}>0$, and the normalized truncated state is well defined.

By Lemma~\ref{lem1}, $\rho_{\rm ext}$ commutes with $P_{<K}$;
hence $\rho_{\rm ext}$ is block diagonal with respect to the retained subspace and its orthogonal complement.
When $0<\delta_K<1$, let
\[
 \rho_{\rm drop}
 =\frac{(I-P_{<K})\rho_{\rm ext}(I-P_{<K})}{\delta_K}.
\]
This operator is positive semidefinite with trace $1$, is orthogonal in support to $\widetilde\rho$, and
\[
 \rho_{\rm ext}=(1-\delta_K)\widetilde\rho+\delta_K\rho_{\rm drop}.
\]
Since the supports of $\widetilde\rho$ and $\rho_{\rm drop}$ are orthogonal,
the positive and negative parts of the difference operator
$\rho_{\rm ext}-\widetilde\rho
=\delta_K(\rho_{\rm drop}-\widetilde\rho)$
are $\delta_K\rho_{\rm drop}$ and $\delta_K\widetilde\rho$, respectively.
Its trace norm is the sum of the traces of these two parts, $2\delta_K$, i.e.,
$\|\rho_{\rm ext}-\widetilde\rho\|_1/2=\delta_K$.

The network unitary transformation preserves the trace norm.
For the operators $M_o\geq0$, $\sum_oM_o=I$, of the complete output photon-number measurement,
decompose the transformed self-adjoint difference operator as $D=D_+-D_-$,
where $D_\pm\geq0$ and $D_+D_-=0$.
Since $\operatorname{tr}(M_oD_\pm)\geq0$, we have
\[
 \sum_o|\operatorname{tr}(M_oD)|
 \leq\sum_o\bigl(\operatorname{tr}(M_oD_+)+\operatorname{tr}(M_oD_-)\bigr)
 =\operatorname{tr} D_++\operatorname{tr} D_-=\|D\|_1.
\]
Dividing both sides by $2$ gives the first inequality in Eq.~\eqref{eq6}.
When $\delta_K=0$, the original state coincides with the truncated state, and the conclusion still holds.
\end{proof}

\begin{proof}[\normalfont\bfseries Proof of Theorem~\ref{thm4}]
Let $B_1,\ldots,B_m$ be the local-identity classes, $\nu_g=|B_g|$,
let $\zeta_g$ be the common delay function of the $g$-th class, and write
$u_{gh}=\operatorname{ord}_0(\zeta_g-\zeta_h)$ $(g\ne h)$.
Let $q_j=\lambda'_j$, $1\leq j\leq\lambda_1$.
If two labels from the same identity class are placed in the same column, then the column-difference product is identically zero.
Thus a column assignment with finite cost corresponds exactly to a matrix satisfying
\[
 a_{gj}\in\{0,1\},\qquad \sum_j a_{gj}=\nu_g,\qquad
 \sum_g a_{gj}=q_j,
\]
where $a_{gj}$ indicates whether the $j$-th column selects a label from the $g$-th class.
Conversely, assigning the labels of each class one-to-one to the columns selected by that class, and then arbitrarily permuting the labels within each column, recovers a tableau.
When $m=1$, only $\lambda=(n)$ is feasible, and then $H_\lambda=0$;
for all other shapes, $H_\lambda=+\infty$. Assume $m\geq2$. For any integer $r\geq1$, the condition $u_{gh}\geq r$ is equivalent to the Taylor coefficients of $\zeta_g$ and $\zeta_h$ at
$\varepsilon^0,\ldots,\varepsilon^{r-1}$ being exactly the same.
As $r$ increases, the classes defined by these coefficients only continue to split and never merge again.
This hierarchical structure can be represented by a tree.
Strictly speaking, from $\zeta_g-\zeta_h=(\zeta_g-\zeta_\ell)+(\zeta_\ell-\zeta_h)$,
the sum of two series cannot produce a new term of order lower than the minimum of their lowest orders; hence
$u_{gh}\geq\min\{u_{g\ell},u_{\ell h}\}$ holds for pairwise distinct $g,h,\ell$.
Construct a tree with root $C_0=\{1,\ldots,m\}$:
for a node $C$ containing at least two class indices, let
$w_C=\min_{g,h\in C,\,g\ne h}u_{gh}$,
and partition it into child nodes according to $g\sim h$ if and only if $g=h$ or $u_{gh}>w_C$.
The above order inequality ensures that this relation is an equivalence relation.
Every nonleaf node has at least two child nodes, and the final leaves are single class indices.
Let $P(C)$ denote the parent node, and set
\[
 \delta_{C_0}=w_{C_0},\qquad
 \delta_C=w_C-w_{P(C)}\quad(C\ne C_0\text{ nonleaf}).
\]
These increments are nonnegative. The least common ancestor means the smallest tree node that contains both leaf indices.
If $g,h$ are first separated into different child nodes under node $C$,
then by the grouping rule $u_{gh}=w_C$; this is exactly the height of their least common ancestor.
Summing the increments along the path from the root to this node, the heights of intermediate nodes cancel term by term, so
$u_{gh}=\sum_{C\supseteq\{g,h\}}\delta_C$,
where the sum is only over nonleaf nodes.
Let $k_{Cj}=\sum_{g\in C}a_{gj}$.
Applying this identity to the unordered pairs of labels in each column and interchanging the order of summation gives
\begin{equation}
 H_\lambda=\min_{(a_{gj})}
       \sum_{C\text{ nonleaf}}\delta_C
                    \sum_{j=1}^{\lambda_1}\binom{k_{Cj}}2.
 \label{eq8}
\end{equation}
Here the matrix satisfies the above $0$--$1$ value conditions and row and column sum constraints;
when no feasible matrix exists, the minimum is $+\infty$.

Next we transform the column-assignment problem into a minimum-cost flow problem.
A flow network is a directed graph whose arcs have capacities and per-unit flow costs, and in which, except at the source and sink, the inflow equals the outflow at every node.
The capacities give the upper bounds on the flow in each arc, and the total cost is the sum of the products of the arc flows and their corresponding per-unit costs.
In this paper, one integer flow unit corresponds to one photon label, and a flow of $n$ means that all labels have been assigned.
Denote the source and sink by $s_0,t_0$, respectively, to distinguish them from the spectral-width parameter $s$.
Create a node $L_g$ for each class and add an arc $s_0\to L_g$,
with capacity $\nu_g$ and cost $0$.
For each nonleaf node $C$ and each column $j$, create a node $C_j$.
Add an arc from $L_g$ to $P(g)_j$ for each column $j$, where $P(g)$ is the parent of leaf $g$, with capacity $1$ and cost $0$.
This ensures that each class appears at most once in each column.
For each nonroot nonleaf node $C$ and each column $j$,
add $q_j$ parallel arcs of unit capacity from $C_j$ to $P(C)_j$,
with costs $0,\delta_C,\ldots,(q_j-1)\delta_C$, respectively.
For the root node, also add $q_j$ parallel arcs of unit capacity from its $j$-th column to $t_0$,
with costs $0,\delta_{C_0},\ldots,(q_j-1)\delta_{C_0}$, respectively.
Find a minimum-cost flow of value $n$ in this network.

Integer capacities and integer supplies and demands guarantee that an integer optimal flow exists whenever a feasible flow exists; see \cite{amo}.
The total capacities at the source and sink are both $n$;
hence an integer flow of value $n$ makes each class select exactly $\nu_g$ distinct columns and each column receive exactly $q_j$ labels.
The arcs in the tree preserve the column index, so the outflow from $C_j$ is $k_{Cj}$.
When sending $k$ units of flow, choosing the $k$ parallel arcs of lowest cost gives cost exactly
$\delta_C(0+1+\cdots+(k-1))=\delta_C\binom{k}2$.
If a flow does not choose the $k$ lowest-cost parallel arcs in a group,
one can replace them by these arcs without changing flow conservation at any node and without increasing the cost;
hence an optimal flow can always be chosen in this way.
Conversely, the label assignment determined by any feasible matrix can be converted into a flow sent upward along the tree;
choosing the lowest-cost arcs in each group of parallel arcs gives an integer flow of the same cost.
Therefore the flow problem has the same optimal value and feasibility as Eq.~\eqref{eq8}.
By Theorem~\ref{thm2}, infeasibility corresponds exactly to identically zero weight,
i.e., $\lambda$ does not dominate the nonincreasing rearrangement of the identity-class sizes.
Let $r_{\rm int}\leq m-1$ be the number of nonleaf nodes of the tree.
The number of nodes and arcs in the constructed network are
\[
 2+m+r_{\rm int}\lambda_1,\qquad
 m+m\lambda_1+r_{\rm int}n.
\]
Here each nonleaf node contributes $\sum_jq_j=n$ parallel arcs,
and the arcs from class nodes to the columns total $m\lambda_1$.
Since $m,\lambda_1\leq n$, both the number of nodes and the number of arcs are $O(n^2)$.
The tree can be constructed in polynomial time by comparing and grouping the order tables.
Let $B$ be the maximum binary bit length of the input orders.
Each arc cost is of the form $k\delta_C$, where $0\leq k<n$,
and $\delta_C$ does not exceed the maximum input order,
so the bit length of the costs is at most $B+O(\log n)$.
Standard polynomial-time algorithms for minimum-cost flow (see \cite{amo}) therefore yield the stated complexity.
The above complexity conclusion does not include the computation of deciding local identity relations or extracting vanishing orders from a black-box representation of arbitrary analytic functions,
nor does it imply that the numerical values or leading coefficients of immanants can be computed in polynomial time.
\end{proof}

\section{Further Discussion}\label{sec4}
Time delays affect multiphoton synchronization and also provide an experimental means of tuning distinguishability.
The results of this paper show that, for independent pure-state photons with a common Gaussian spectral envelope,
the weights of the non-fully-symmetric components are suppressed in the near-synchronization regime according to a definite power law.
Improving the time matching within groups first and then tuning the inter-group delays can produce a different evolution law for the weights than scaling all delay differences down by the same factor.
These weights are determined by the input wave packets, and their influence on the output photon-number distribution also depends on the network and on the normalized spatial states of the respective components.

The spectral width and the synchronization accuracy jointly determine the dimensionless mismatch parameter
$S=s^2\sum_i(\theta_i-\bar\theta)^2$.
The same arrival-time deviation corresponds to a larger $S$ under a broader spectrum.
When $\kappa_\lambda(\boldsymbol{y})>0$, dividing Eq.~\eqref{eq3} by $\kappa_\lambda(\boldsymbol{y})$ gives
\[
 0\leq1-\frac{p_\lambda(\boldsymbol{y})}{\kappa_\lambda(\boldsymbol{y})}
 \leq1-e^{-S}\leq S.
\]
Therefore, when $\kappa_\lambda$ is used to approximate the weight of an exchange-symmetry component, the upper bound on the deviation relative to this leading-order approximation depends only on $S$,
and it is not necessary to assume separately that the pairwise delays have a positive minimum separation.
Combining this with the output error bound in Theorem~\ref{thm3}, one can also convert an allowed statistical deviation into a sufficient condition on the time-control accuracy.

From the projection formula for the fully symmetric component in Section~\ref{sec2},
after retaining only this component and normalizing it, one obtains the output distribution of ideally indistinguishable photons.
As the number of retained types increases, the trace distance between the truncated state and the original state does not increase, because
\[
 \delta_K-\delta_{K+1}
 =\sum_{\lambda:\,b_\lambda=K}p_\lambda(\boldsymbol{y})\geq0.
\]
The Poisson-tail probability bound in Eq.~\eqref{eq5} also does not increase as $K$ increases,
and the difference between two adjacent terms is $e^{-S}S^K/K!$.
These conclusions control the trace distance of the truncated state and the uniform output error bound,
but they do not guarantee that, for a fixed network, the total-variation distance between the output distributions decreases monotonically with $K$.
Moreover, when $S>0$, the ratio of two adjacent terms in the simplified bound $S^K/K!$ is $S/(K+1)$.
Thus the simplified bound need not be nonincreasing in $K$; it decreases whenever $K+1>S$.
The corresponding error bound is independent of the network structure and covers multiphoton events within the same output channel.
If one postselects and normalizes on a subset of counting events, their occurrence probability must also be taken into account.

HOM interference of photon pairs can be used to calibrate the time matching.
In experiments, one must also assess the statistical error of finite sampling, measurement noise, and delay-control error.
For general light sources, multiphoton interference may also depend on cyclic product phases that cannot be determined from the magnitudes of pairwise internal-state overlaps,
see \cite{menssen}.
Spectral differences, mixed or correlated input states, frequency-dependent networks, and time-resolved detection
are physical factors that need to be included in further studies of synchronization errors.

\noindent{\bf Data Availability}\\
{No data were used to support this study.}

\end{document}